\documentclass[preprint2]{aastex}

\slugcomment{}

\shorttitle{Orientation effects in Bipolars}
\shortauthors{Schwarz et al.}

\begin{document}


\title{Orientation effects in Bipolar nebulae}


\author{Hugo E. Schwarz$^1$, Hektor Monteiro$^{1,2}$, \& Ryan Peterson$^{3}$}

\affil{1 Cerro Tololo Inter-American Observatory\altaffilmark{1}, 
Casilla 603, Colina El Pino S/N, La Serena, Chile}

\affil{2 Department of Physics and Astronomy, Georgia State University,  
1, Park Place South, Atlanta, GA 30302, USA}

\affil{3 Lawrence University, Appleton, Wisconsin, USA\altaffilmark{3}}

\altaffiltext{1}{Cerro Tololo Inter-American Observatory, National Optical 
Astronomy Observatory, operated by the Association of Universities for
Research in Astronomy, Inc., under a cooperative agreement with the
National Science Foundation.}
\altaffiltext{3}{2003 CTIO/NOAO/AURA REU Program student}



\begin{abstract}

We show that the inclination to the line of sight of bipolar nebulae
strongly affects some of their observed properties. We model these
objects as having a spherically symmetric Planetary Nebula and a dusty
equatorial density enhancement that produces extinction that varies
with the viewing angle. Our sample of 29 nebulae taken from the
literature shows a clear correlation between the inclination angle and
the near-IR and optical photometric properties as well as the apparent
luminosity of the objects. As the inclination angle increases --the
viewing angle is closer to the equatorial plane-- the objects become
redder, their average apparent luminosity decreases, and their average
projected expansion velocity becomes smaller.

We compute two-dimensional models of stars embedded in dusty disks of
various shapes and compositions and show that the observed data can be
reproduced by disk-star combinations with reasonable parameters. To
compare with the observational data, we generate sets of model data by
randomly varying the star and disk parameters within a physically
meaningful range.

We conclude that a only a smooth pole to equator density gradient
agrees with the observed phenomena and that thin, equatorially
concentrated disks can be discarded.

\end{abstract}




\section{Introduction}

Planetary Nebulae (PNe) and Symbiotic Nebulae are the visible remains
of a heavy mass loss period near the end of the lives of most low and
intermediate mass stars (\cite{G97}). A fraction of PNe is bipolar
(BPNe) and this morphology is likely caused by having a binary for
central star (\cite{S04}). BPNe have special properties (\cite{CS95})
and are important objects for the study of outflows, mass loss, and
binarity in PNe Several meetings have been dedicated to asymmetrical
PNe (\cite{HS95}, \cite{KSR00}, \cite{MKBS04}), see references herein.

The orientation angle on the plane of the sky of asymmetrical PNe has
been studied more than 30 years ago by \cite{MH75} and more recently
by \cite{P97} who both found an apparent alignment of the long axes of
PNe with the plane of the Galaxy. The paper by \cite{CAM98} --based on
a larger sample and using more rigorous statistical methods-- showed
that there is no significant alignment and that PNe are essentially
randomly distributed on the sky. Assuming that this random
distribution also holds for the inclination with respect to the line
of sight --where the inclination, i, is taken to be 0$^o$ for a
pole-on nebula and 90$^o$ for an object viewed with its main axis
parallel to the plane of the sky--we investigate the possible effects
of orientation on the observational properties PNe.
 
\cite{SHK01} studied the observational orientation effects in six
bipolar Proto-PNe and found that objects are redder when viewed
edge-on i.e. have the inclination angle i near 90$^o$, as expected
from simple geometrical extinction considerations nd assuming some
equatorial density enhancement. From these data they concluded that
Proto-PNe generally have bipolar shapes and asymmetrical dust disks.

Here we present a more general look at the phenomenon and apply a
simple model to a larger sample of objects that we know to be bipolar,
to see if we observe differences as a function of inclination
angle. We study a range of overall geometries and dust properties of
the equatorial tori, different black-body temperature ratios between
the dust and central stars, and simulate observational data sets by
letting these model parameters vary randomly over a restricted and
physically meaningful range. These data sets we then compare directly
with our observed results.

\section{Observational material}

We have collected a sample of 29 bipolar planetary nebulae and
symbiotic nebulae that had sufficient data available in the literature
to construct a reasonable spectral energy distribution (SED), with at
least some data points in the optical, near-IR, and mid-IR wavelength
ranges. Table\,1 lists our sample of objects and their calculated
relative luminosities in each of the three bands that we defined as
follows.

\begin{table*}

\caption{Our observed sample of BPNe. Listed are: PK name, common name, 
B, V, R, J, H, and K in magnitudes; 12, 25, 60, 100 fluxes in Janskys,
the inclination angle (i) in degrees, and the apparent luminosity (L) in
W.m$^2$ scaled as the square of the distance as derived from the SED.}

\begin{tabular}{llcccccccccccc}
\hline\hline
{Name} & {B} & {V} & {R} & {J} & {H} & {K} & {12} & {25} & {60} & {100} & {i} & {L} \\
\hline
 19W32      &  18.2  &  17.2 &  14.3 &     10.8 &  9.3  &  8.6  &      3.4   &  12.6  &  23.0  &  356.0 &     80. &    1.99244e-12\\
 Hb5        &  10.3  &  8.2  &   -   &     9.5  &  8.8  &  8.6  &      11.68 &  79.24 &  134.50&  311.8 &     65. &    1.39518e-11\\
 He2-25     &  16.4  &   -   &  13.9 &     10.7 &  9.7  &  9.5  &      0.742 &  2.11  &  3.39  &  41.0  &     90. &    5.23481e-13\\
 He2-36     &  11.9  &  11.2 &  8.9  &     9.8  &  9.6  &  9.4  &      0.42  &  4.88  &  6.40  &  8.8   &     30. &    1.67762e-12\\
 He2-111    &  16.3  &  16.7 &   -   &     13.5 &  11.1 &  9.7  &      1.28  &  3.08  &  11.21 &  117.2 &     70. &    7.42040e-13\\
 He2-114    &  18.7  &   -   &  16.3 &      -   &   -   &   -   &      2.0   &  0.38  &  3.08  &  38.46 &     45. &    3.06317e-13\\
 He2-145    &  19.1  &   -   &  16.4 &     12.0 &  10.9 &  10.1 &      3.1   &  2.67  &  44.75 &  201.0 &     55. &    2.21225e-12\\
 IC4406     &  12.7  &   -   &  11.4 &      -   &   -   &  10.1 &      0.33  &  2.69  &  21.08 &  25.35 &     65. &    1.18293e-12\\
 07131-0147 &  15.4  &   -   &  11.7 &     10.2 &  9.5  &  9.1  &      2.59  &  4.22  &  3.96  &  3.68  &     60. &    8.54946e-13\\
 K3-46      &  19.1  &   -   &  16.3 &     13.8 &  13.0 &  12.7 &      0.7   &  0.25  &  0.73  &  5.87  &     65. &    1.04686e-13\\
 M1-8       &  11.7  &   -   &  10.4 &     10.7 &  10.5 &  9.8  &      0.25  &  0.66  &  3.10  &  5.54  &     25. &    7.55110e-13\\
 M1-13      &  11.0  &   -   &  9.9  &      -   &   -   &   -   &      0.25  &  0.70  &  4.51  &  9.37  &     25. &    4.58175e-12\\
 M1-16      &  12.7  &   -   &  10.0 &     13.0 &  13.2 &  12.2 &      0.32  &  2.33  &  9.45  &  7.59  &     60. &    7.58743e-13\\
 M1-28      &  18.9  &   -   &  16.0 &     15.1 &  14.2 &  13.8 &      2.4   &  0.41  &  2.77  &  11.75 &     55. &    3.11362e-13\\
 M1-91      &  16.3  &  15.0 &  12.4 &      -   &   -   &   -   &      3.85  &  8.29  &  12.14 &  9.56  &     75. &    1.26640e-12\\
 M2-9       &  11.3  &   -   &  9.5  &     10.9 &  9.2  &  7.0  &      50.50 &  110.20&  123.60&  75.84 &     75. &    1.19351e-11\\
 M2-48      &  15.2  &   -   &  10.0 &     15.9 &  14.4 &  13.9 &      1.03  &  0.76  &  7.42  &  35.86 &     45. &    7.52791e-13\\
 M3-28      &  15.6  &   -   &  10.3 &      -   &  9.6  &  8.0  &      3.53  &  2.56  &  46.94 &  320.60&     60. &    3.16344e-12\\
 MyCn18     &  11.8  &   -   &  9.6  &     11.8 &  10.3 &  9.8  &      1.80  &  20.66 &  24.28 &  13.24 &     55. &    2.17540e-12\\
 Mz1        &  12.8  &   -   &  11.6 &      -   &   -   &   -   &      0.48  &  1.40  &  14.45 &  92.96 &     50. &    1.57525e-12\\
 Mz3        &  10.8  &   -   &  8.0  &     9.2  &  7.4  &  5.6  &      88.76 &  343.20&  277.00&  112.60&     65. &    3.00787e-11\\
 NGC650-51  &   -    &  17.7 &   -   &     16.0 &  15.9 &  15.6 &      0.28  &  2.79  &  6.80  &  9.29  &     60. &    3.09241e-13\\
 NGC2346    &  11.5  &  11.2 &   -   &     10.2 &  9.4  &  8.4  &      0.47  &  0.88  &  7.97  &  13.40 &     45. &    1.15817e-12\\
 NGC2440    &  11.8  &   -   &  11.0 &     10.6 &  10.8 &  10.1 &      3.59  &  28.01 &  43.47 &  26.30 &     65. &    2.64962e-12\\
 NGC2818    &  16.8  &   -   &  16.8 &      -   &   -   &  18.2 &      0.32  &  1.00  &  2.30  &  2.89  &     90. &    1.25058e-13\\
 NGC2899    &  17.8  &   -   &  16.3 &      -   &   -   &   -   &      0.27  &  1.42  &  5.13  &  10.91 &     75. &    2.59349e-13\\
 NGC6072    &  11.6  &   -   &  10.1 &      -   &   -   &   -   &      0.38  &  2.87  &  24.89 &  31.17 &     45. &    4.33377e-12\\
 NGC6302    &   -    &  9.1  &  7.1  &     9.5  &  9.7  &  8.6  &      32.08 &  335.90&  849.70&  537.40&     70. &    3.66894e-11\\
 NGC6445    &  11.6  &   -   &  12.5 &     10.9 &  9.6  &  9.2  &      1.50  &  15.01 &  44.44 &  43.23 &     50. &    2.22741e-12\\
 NGC6537    &  11.0  &   -   &  12.0 &     10.3 &  10.4 &  9.3  &      7.72  &  58.30 &  189.90&  166.10&     60. &    7.70877e-12\\
 NGC7026    &  15.3  &  14.2 &  11.7 &     10.2 &  11.0 &  9.6  &      2.39  &  18.35 &  42.74 &  30.91 &     50. &    2.03701e-12\\
 Na2        &  18.7  &   -   &   -   &     15.9 &  15.7 &  14.6 &      0.46  &  0.33  &  0.93  &  1.32  &      -  &    7.45171e-14\\
 Sa2-237    &  16.1  &  15.5 &  15.0 &     13.1 &  12.6 &  11.8 &      1.17  &  6.01  &  16.56 &  8.29  &     70. &    6.57414e-13\\
 SH1-89     &  15.8  &   -   &  13.6 &     14.4 &  13.2 &  12.8 &      0.89  &  0.25  &  2.14  &  20.15 &      -  &    1.88058e-13\\
 TH2-B      &  19.4  &   -   &  16.7 &      -   &   -   &   -   &      1.40  &  2.15  &  3.22  &  34.68 &      -  &    3.24659e-13\\
 He2-104    &  13.5  &  14.6 &  11.1 &     10.8 &  8.9  &  7.0  &      8.56  &  9.09  &  6.83  &  9.77  &     50. &    2.14718e-12\\
 BI Cru     &  12.5  &  10.9 &  11.3 &     7.3  &  6.0  &  4.8  &      17.29 &  15.34 &  11.84 &  119.10&     40. &    1.20309e-11\\
 R Aqr      &  10.0  &  6.2  &   -   &     0.5  & -0.1  & -0.8  &      1577.0&  543.80&  66.65 &  16.60 &     20. &    2.83343e-09\\
 AS201      &  12.4  &  12.5 &  11.3 &     10.6 &  10.2 &  9.5  &      4.12  &  38.33 &  76.14 &  41.49 &      -  &    3.59771e-12\\         
\hline
\end{tabular}
\end{table*}

After constructing the SEDs we defined three different apparent
luminosities: visible (BVR), NIR (JHK) and IRAS (12, 25, 60, \&
100\,$\mu$m) and divided these three luminosities by the total
(near-bolometric) apparent luminosity as derived from the SED. All
luminosities were calculated by integrating the F($\lambda$) curve
over the appropriate wavelength range. Possible differences in the
line of sight interstellar extinctions were ignored. These differences
would affect mainly the BVR luminosity since the extinction is a
strongly declining function of wavelength. For those objects that
could have their absolute luminosity determined, we also investigated
the dependence of the absolute luminosity on the inclination
angle. There were only 7 objects with good enough distances for
computing an believable absolute luminosity.

The last column in Table\,1 shows the average of the three estimated
inclination angles for each object. The inclination angles were
determined by the three authors independently, to obtain an idea of
the associated uncertainties in deriving inclinations from optical
images. The process has a subjective component but we found (somewhat
to our surprise!) that for nearly all cases our independent estimates
were in agreement to within 10\,$^o$ or so, accurate enough for the
purposes of this paper. For only three ``recalcitrant'' objects did we
differ by as many as 30\,$^o$ in our estimates.Typical standard deviations
of the mean of the three estimates are 5-8\,$^o$ and in all cases they
were below 18\,$^o$.

In Figure\,1 we plot the relative luminosities for each band --visible
(BVR); NIR (JHK); MIR (IRAS bands)-- as a function of the inclination
angle. One can clearly see the increase of relative MIR as well as the
decrease of JHK and BVR with increasing inclination for the sample.

\begin{figure*}[!ht]
\includegraphics[scale=0.70]{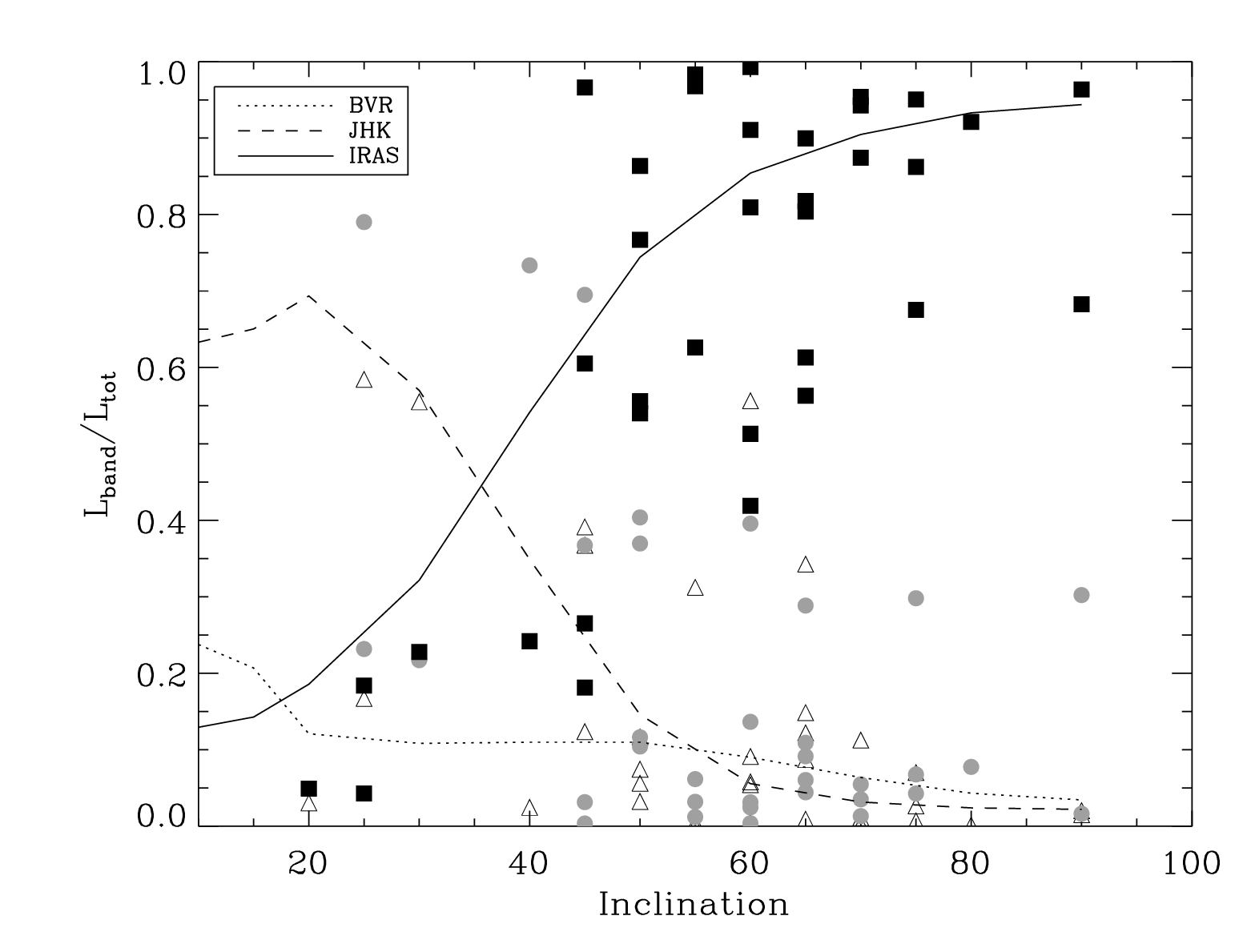}
\caption{The observed relative luminosities in the BVR (open triangles), 
JHK (grey dots), and IRAS (black squares) bands as a function of the 
inclination angle. The best fitting model is overplotted as lines (solid = IRAS;
dashed = JHK; dotted = BVR).}
\end{figure*}

\begin{table*}

\caption{Apparent total, BVR, JHL, and IRAS luminosities and inclinations 
for objects studied. See text for the detailed  definition of these 
luminosities.}

\begin{tabular}{llccccc}
\hline\hline
{PK} & {Name} & {BVR} & {JHK} & {IRAS} & {$L_{Total}$} & {Incl.}\\
\hline

     -     &  07131-0147 & $ 5.65\times 10^{-5}$  &  $5.64\times 10^{-2}$  &  $2.65\times 10^{-1}$  &  $6.78\times 10^{-1}$  &    60    \\ 
359.2+01.2 &  19W32      & $ 1.81\times 10^{-4}$  &  $1.08\times 10^{-3}$  &  $9.48\times 10^{-2}$  &  $9.04\times 10^{-1}$  &    80    \\  
     -     &  BI Cru     & $ 8.51\times 10^{-4}$  &  $1.61\times 10^{-2}$  &  $7.35\times 10^{-1}$  &  $2.49\times 10^{-1}$  &    40    \\  
359.3-00.9 &  Hb5        & $ 7.45\times 10^{-4}$  &  $1.70\times 10^{-1}$  &  $4.70\times 10^{-2}$  &  $7.83\times 10^{-1}$  &    65    \\
004.8+02.0 &  He2-25     & $ 5.99\times 10^{-5}$  &  $9.33\times 10^{-3}$  &  $2.34\times 10^{-1}$  &  $7.57\times 10^{-1}$  &    90    \\
359.6-04.8 &  He2-36     & $ 1.14\times 10^{-4}$  &  $4.10\times 10^{-1}$  &  $1.88\times 10^{-1}$  &  $4.01\times 10^{-1}$  &    30    \\ 
315.4+09.4 &  He2-104    & $ 2.05\times 10^{-4}$  &  $2.33\times 10^{-2}$  &  $5.07\times 10^{-1}$  &  $4.70\times 10^{-1}$  &    50    \\  
315.0-00.3 &  He2-111    & $ 5.64\times 10^{-5}$  &  $1.22\times 10^{-3}$  &  $5.83\times 10^{-2}$  &  $9.40\times 10^{-1}$  &    70    \\ 
318.3-02.0 &  He2-114    & $ 2.49\times 10^{-5}$  &  $1.75\times 10^{-3}$  &  $2.53\times 10^{-2}$  &  $9.73\times 10^{-1}$  &    45    \\
331.4+00.5 &  He2-145    & $ 1.14\times 10^{-4}$  &  $4.13\times 10^{-4}$  &  $3.24\times 10^{-2}$  &  $9.67\times 10^{-1}$  &    55    \\ 
319.6+15.7 &  IC4406     & $ 5.12\times 10^{-5}$  &  $9.05\times 10^{-2}$  &  $9.98\times 10^{-2}$  &  $8.10\times 10^{-1}$  &    65    \\
069.2+03.8 &  K3-46      & $ 8.12\times 10^{-6}$  &  $6.44\times 10^{-3}$  &  $7.91\times 10^{-2}$  &  $9.14\times 10^{-1}$  &    65    \\ 
210.3+01.9 &  M1-8       & $ 4.13\times 10^{-5}$  &  $3.93\times 10^{-1}$  &  $3.30\times 10^{-1}$  &  $2.77\times 10^{-1}$  &    25    \\
232.4-01.8 &  M1-13      & $ 6.57\times 10^{-5}$  &  $4.07\times 10^{-1}$  &  $3.68\times 10^{-1}$  &  $2.25\times 10^{-1}$  &    25    \\ 
226.7+05.6 &  M1-16      & $ 6.01\times 10^{-5}$  &  $3.22\times 10^{-1}$  &  $1.01\times 10^{-1}$  &  $5.77\times 10^{-1}$  &    60    \\
0.060+03.1 &  M1-28      & $ 2.11\times 10^{-5}$  &  $2.91\times 10^{-3}$  &  $3.11\times 10^{-2}$  &  $9.66\times 10^{-1}$  &    55    \\ 
      -    &  M1-91      & $ 8.78\times 10^{-5}$  &  $1.62\times 10^{-2}$  &  $7.41\times 10^{-2}$  &  $9.10\times 10^{-1}$  &    75    \\
010.8+18.0 &  M2-9       & $ 1.28\times 10^{-3}$  &  $2.43\times 10^{-2}$  &  $2.02\times 10^{-1}$  &  $7.74\times 10^{-1}$  &    75    \\
062.4-00.2 &  M2-48      & $ 5.02\times 10^{-5}$  &  $3.04\times 10^{-1}$  &  $2.68\times 10^{-2}$  &  $6.69\times 10^{-1}$  &    45    \\ 
021.8-00.4 &  M3-28      & $ 1.87\times 10^{-4}$  &  $4.40\times 10^{-2}$  &  $1.30\times 10^{-1}$  &  $8.26\times 10^{-1}$  &    60    \\
307.5-04.9 &  MyCn18     & $ 1.58\times 10^{-4}$  &  $1.42\times 10^{-1}$  &  $4.47\times 10^{-2}$  &  $8.13\times 10^{-1}$  &    55    \\ 
322.4-02.6 &  Mz1        & $ 5.12\times 10^{-5}$  &  $7.30\times 10^{-2}$  &  $7.18\times 10^{-2}$  &  $8.55\times 10^{-1}$  &    50    \\
331.7-01.0 &  Mz3        & $ 2.90\times 10^{-3}$  &  $3.29\times 10^{-2}$  &  $1.68\times 10^{-1}$  &  $7.99\times 10^{-1}$  &    65    \\   
130.9-10.5 &  NGC650-51  & $ 1.32\times 10^{-4}$  &  $2.11\times 10^{-4}$  &  $5.84\times 10^{-1}$  &  $4.16\times 10^{-1}$  &    60    \\   
215.6+03.6 &  NGC2346    & $ 6.03\times 10^{-5}$  &  $2.07\times 10^{-1}$  &  $4.22\times 10^{-1}$  &  $3.71\times 10^{-1}$  &    45    \\  
234.8+02.4 &  NGC2440    & $ 2.28\times 10^{-4}$  &  $4.24\times 10^{-2}$  &  $7.88\times 10^{-2}$  &  $8.79\times 10^{-1}$  &    50    \\  
     -     &  R Aqr      & $ 7.10\times 10^{-2}$  &  $3.67\times 10^{-2}$  &  $8.01\times 10^{-1}$  &  $1.62\times 10^{-1}$  &    20    \\ 
011.1+07.0 &  Sa2-237    & $ 3.00\times 10^{-4}$  &  $7.30\times 10^{-3}$  &  $1.80\times 10^{-2}$  &  $9.60\times 10^{-1}$  &    70    \\  

\hline
\end{tabular}
\end{table*}

The observed relative luminosities in the BVR, JHK, and IRAS bands are
shown as a function of inclination angle in Figure\,1.

\section{Models}

\subsection{A simplified 2-D radiative transfer code}

To try to explain this observed behavior, we propose that these
bipolar systems are composed of a dusty equatorial density enhancement
being irradiated by a star or a binary system. The UV flux from this
system heats up the dust that then re-emits the radiation in the
IR. The combined effect of this re-radiation with the extinction
produced by the density distribution is what produces the observed
effect.

To simulate this system we created a simple 2-D radiative transfer
code using the IDL package. This code works by creating a 2-D grid of
cells, each having a given density. By requiring conservation of
luminosity at each cell and iterating we determine the temperature of
the dust. The temperature is determined from the following equation:

$\int C_{abs}cu_{\lambda }d\lambda =\int C_{abs}4\pi B_{\lambda }(T)d\lambda $

where T is the temperature, $u_{\lambda }$ is the energy density of
the radiation field and $C_{abs}$ is the absorption cross-section,
shown in Figure\,2 as a function of the wavelength. Secondary
radiation is as of yet not accounted for. Scattering is calculated
assuming an isotropic phase function.

Some simplifying assumptions are made concerning the dust grains:

\begin{enumerate}
\item the dust species are well mixed, allowing for the use of average 
cross-sections;
\item the dust mixture is in equilibrium with the radiation field which 
implies a single dust temperature for all dust species;
\item the dust-to-gas ratio is the same as the one for the interstellar medium.
\end{enumerate}

For the dust characteristics we then adopted typical interstellar
grains with R$_v$\,=\,3.1 and absorption cross-section calculated by
\cite{LD01} with a dust to gas ratio of 0.008. The absorption
cross-section is shown in Figure\,2.

\begin{figure}[!ht]
\includegraphics[scale=0.45]{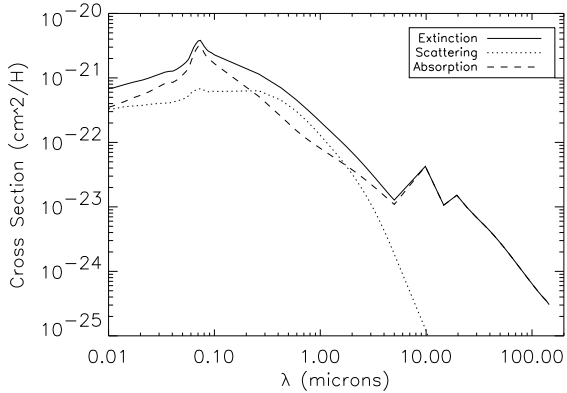}
\caption{Absorption cross-sections used in the model 
calculations. From \cite{LD01}.}
\end{figure}

With the 2-D temperature and density grids, we constructed a 3-D grid
where the calculated intensity emitted by the dust plus scattered
radiation (assumed isotropic) is obtained for each cell. Using this
cube and the extinction cross-section from \cite{LD01} we calculated
the image projected onto the sky for a given line of sight. The total
emitted radiation was obtained from this projected image and was then
compared to observational results.

\subsection{Calculated Models}

Using the simple code described above we initially calculated models
for the three basic disk-like structures shown in Figure\,3, 1st
column. From top to bottom these are: a) a torus; b) a flat disk and
c) a curved disk. A fourth distribution was used with a much more
smoothly varying density gradient from pole to equator, shown as the
last entry in the 1st column of Figure\,3. For simplicity we adopted
the same physical size and peak density for all structures. 

\begin{figure*}[!ht]
\includegraphics[scale=0.90]{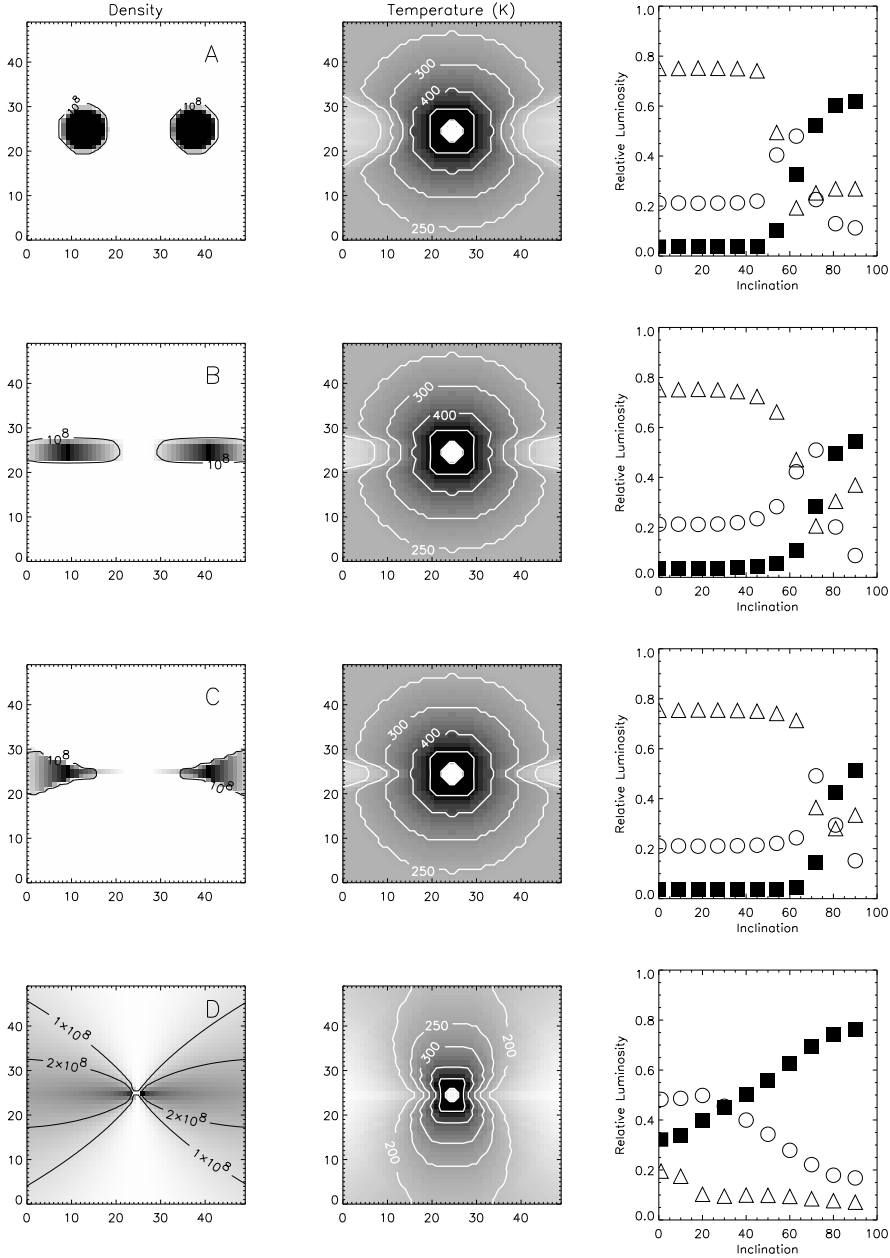}
\caption{Density and temperature maps obtained for disk structure 
as well as its respective relative luminosity correlation plot.}
\end{figure*}

For the 4 disk shapes adopted we obtained the temperature maps shown
in Figure\,3, column 2. These maps show the shadowing effect of the
dense structures. With the calculated SEDs we also constructed plots
with the correlations of the relative luminosities for each of the
structures as a function of inclination angle, also shown in
Figure\,3, right most column. These can then be compared to the plots
of Figure\,1 which shows the observed behavior.

It is not trivial to compare the model result with the much
``noisier'' observational plot. The noise, or spread is likely caused
by stellar and dust properties varying from object to object and also
by the errors on estimating the inclination angles for each individual
nebula.  To make the comparison easier, we generated sets of model
data by allowing the stellar and dust properties to vary randomly
within a physically meaningful range of values. This simulates
observing a large sample of ``real'' objects with difference
parameters as was done in the observed sample, allowing a direct
comparison of the model and observational samples.

\begin{figure*}[!ht]

\includegraphics[scale=0.8]{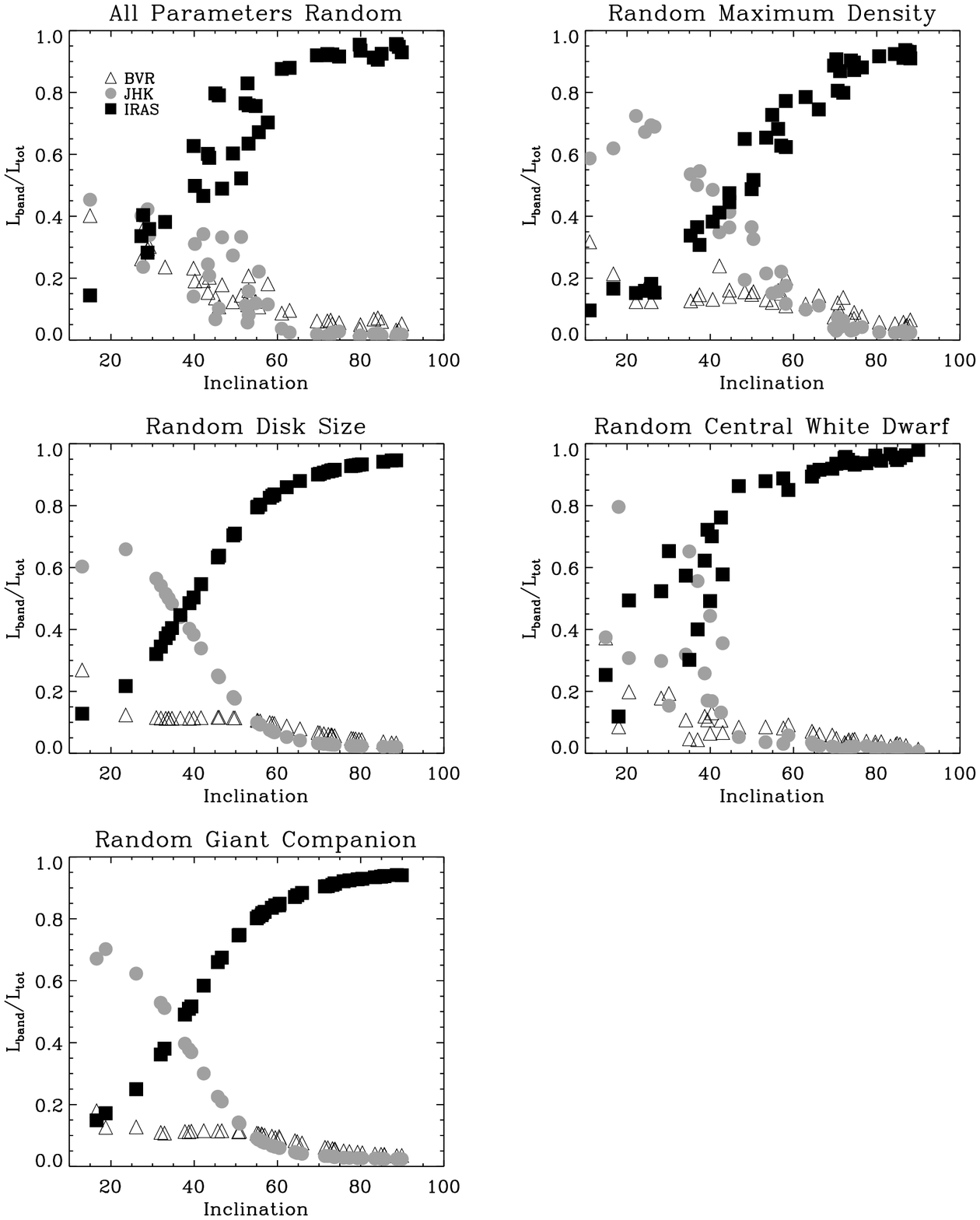}

\caption{Model generated data sets for comparison with observed data. 
The parameters are allowed to vary randomly over a range defined in the 
text. The five graphs show: all parameters varying, only the disk size 
varying, the maximum density, the giant companion, and the white dwarf, 
all as indicated in the figure. }
\end{figure*}

To accomplish this we ran a set of 29 models, equivalent to the
observational sample we had, plotted in Figure\,4. For each object the
code was executed with input parameters selected in a random fashion
with boundaries as described below:\\

\noindent
Disk density: 5.0.10$^8$\,cm$^{-3}$; radius:  4.0.10$^{15}$\,cm\\ 
White Dwarf: T$_{Eff}$\,=\,10$^5$\,K; Luminosity\,=\,500\,L$_{\odot}$\\ 
Giant: T$_{Eff}$\,=\,5.10$^3$\,K; Luminosity\,=\,1000\,L$_{\odot}$\\

The parameters were allowed to vary in the range 0.5 to 1.0 times the
parameter value in a uniform random distribution. The average physical
characteristics of these disks are (in M$_{\odot}$:\\

\noindent
Mass of dust in Disk =    3.64-05\\
Mass of Gas in Disk  =    0.0044\\
Mass of Disk+Shell   =    0.0048\\

The adopted shell has a radius of 10$^{17}$\,cm, a density of
1000\,cm$^{-3}$, and simulates the presence of a generic planetary
nebula in addition to the disk.

\section{Absolute luminosity effects}

Another predicted effect is that high inclination objects should have
lower apparent luminosities because only the equatorial ``donut'' is
seen, while for low inclination objects the central object and donut
are observed, giving an apparent over-luminosity. Clearly, this effect
is only possible in objects with an asymmetrical matter
distribution. Averaging over all angles the total luminosity for a
sample of objects is \\

L$_{tot}$\,=\,n.L$_{ave}$ \\

so that energy is conserved. The excess luminosity of the low-i
objects is exactly compensated by the under-luminous high-i objects.

To check the predicted behavior we generated a sample in the same way
as done for the relative luminosity effect discussed above and plotted
the observed luminosity as a function of inclination angle. The plot
is shown in Figure\,5.

\begin{figure}[!ht]
\includegraphics[scale=0.45]{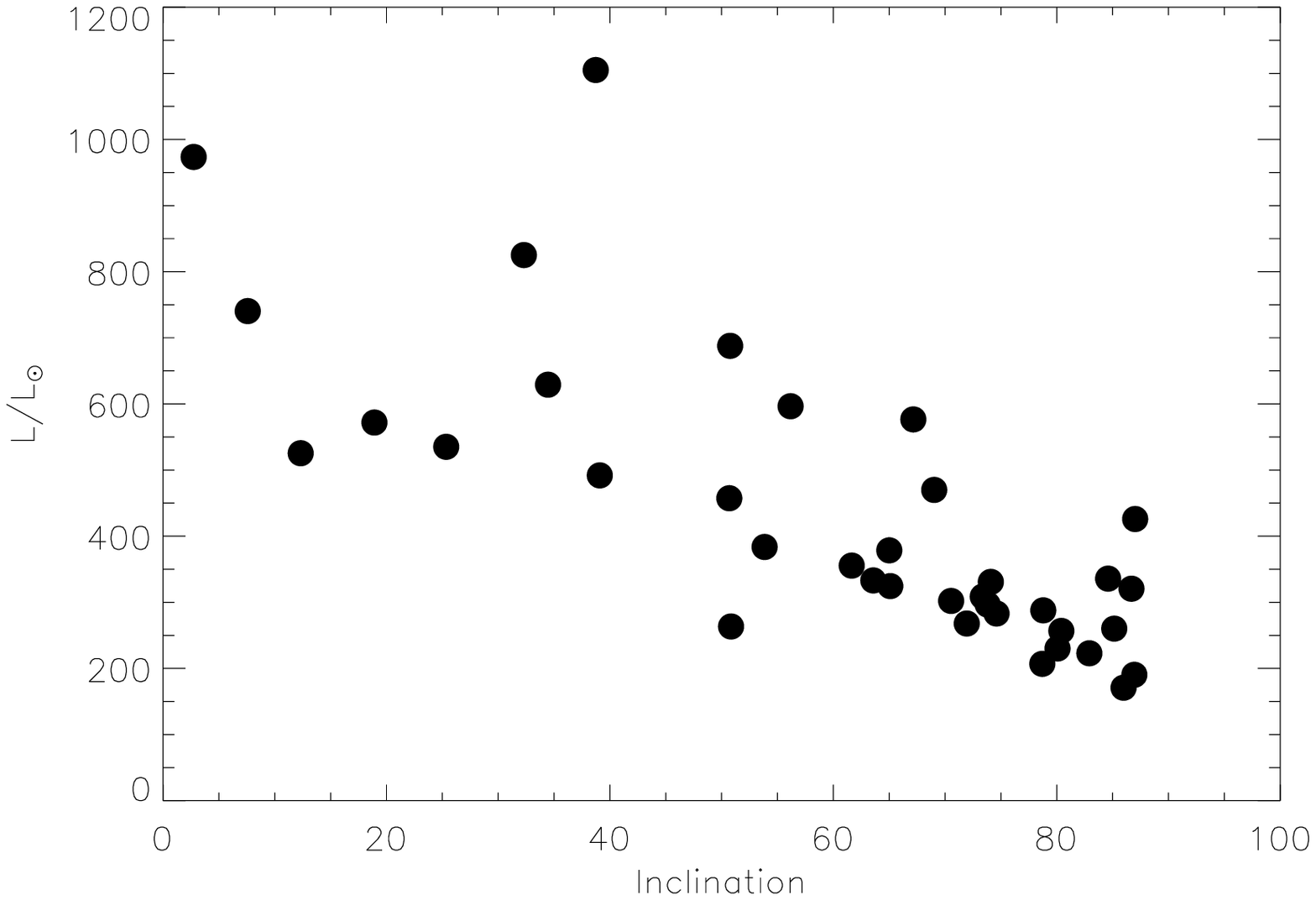}
\caption{Plot of the luminosities of a sample of BPNe as a function 
of inclination angle.}
\end{figure}

For a subset of seven objects we have reliable distances and therefore
were able to determine their absolute luminosities. These objects are
listed in Table\,3 together with their luminosities, distances, and
inclination angles.

\begin{table}

\caption{Objects for which distances are known with their observed 
luminosities. Listed are: Object, luminosity, distance, inclination angle. 
The average luminosities are for objects with i\,$\geq$\,45$^o$ 
L\,=\,346\,L$_{\odot}$ and for i\,$\leq$\,45$^o$ L\,=\,3550\,L$_{\odot}$.}
\begin{tabular}{lccc}
\hline\hline
{Object} & {L(L$_{\odot}$} & d(pc) & i($^o$) \\
\hline
 
 Sa\,2-237 & 340 & 2100 & 70 \\
 M\,2-9    & 553 & 640  & 75 \\
 He\,2-104 & 205 & 800  & 50 \\
 He\,2-111 & 440 & 2800 & 70 \\
 M\,1-16   & 194 & 1800 & 70 \\
 R~Aqr     & 2800& 200  & 20 \\
 BI~Cru    & 4300& 180  & 40 \\
\hline
\end{tabular}
\end{table}

There is a significant difference in the mean luminosities of the two
groups, whereby the high-i group has the lower luminosity, as
predicted. The standard deviation of the high-i group mean is 154 and of
the low-i group 1060. The difference between the mean values is 3204
nearly a 3\,$\sigma$ result.

Another sanity check on the randomness of the angle distribution on
the sky is to count the number of objects in inclination angle bins
and compare these with the theoretically expected numbers which should
go as sin(i) if the distribution on the sky is truly random.  We have
7\% of objects in the 0-30$^o$ bin, 52\% in the 31-60$^o$ bin, and 41\%
in the 61-90$^o$ bin. Noting that there are 5 objects with 60$^o$
inclinations which happen to fall in the 31-60$^o$ bin. Distributing
these equally over this and the next bin, we obtain 7\%, 43\%, and
50\%. Theory predicts --by integrating sin(i) over the same three angle
bins, respectively, 13\%, 37\%, and 50\% of the objects in the bins,
relatively close to the observed values, with or without the
60$^o$ object correction.

\section{Projected expansion velocities}

The expansion velocities of BPNe with low inclinations should -all
other things being equal- be on average higher than those of high
inclinations objects. To test this idea for our sample we used the
published expansion velocities from \cite{CS95}.  To correct as much
as we can for intrinsic differences in expansion velocities between
objects, we took the aspect ratio of an object to be proportional to
its expansion velocity. This makes sense because the polar expansion
compared to the typical PNe expansion of 15\,km.s$^{-1}$ determines
the aspect ratio of the object. Plotting the expansion velocity
divided by aspect ratio against the inclination angle for all those
objects for which we have data, we see in Figure\,6 that there is a
correlation between these parameters. When plotting the expansion
velocity without correcting for the aspect ratio, the correlation is
also there. This is unlikely to be physical and must be related to the
projection of the true expansion velocities. In Table\,4 we list the
objects with their expansion velocities and aspect ratios.

\begin{figure}[ht]
\includegraphics[width=\columnwidth]{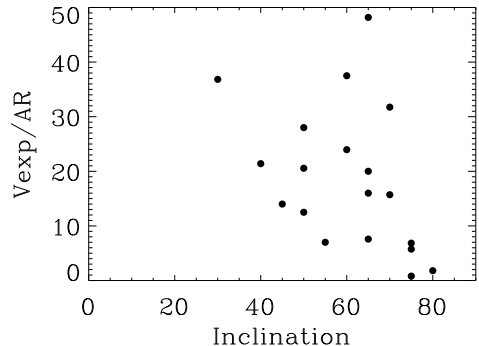}
\caption{Plot of the expansion velocities of our sample as a function
of inclination angle corrected for the objects aspect ratio.}
\end{figure}

\begin{table}[ht]

\caption{Objects for which the expansion velocity is known with their 
aspect ratio.}
\begin{tabular}{lcc}
\hline\hline
{Object} & {$V_{exp}$ ($km/s$)} & {Aspect Ratio (AR)} \\
\hline
19W32    &      9   &      5.0  \\
Hb5      &      106 &      2.2  \\
He2-36   &      70  &      1.9  \\
He2-111  &      127 &      4.0  \\
IC4406   &      25  &      3.3  \\
M1-16    &      127 &      5.3  \\
M1-91    &      4   &      5.0  \\
M2-9     &      69  &      12.0 \\
MyCn18   &      14  &      2.0  \\
Mz3      &      76  &      3.8  \\
NGC2346  &      35  &      2.5  \\
NGC2440  &      32  &      2.0  \\
NGC2899  &      13  &      1.9  \\
NGC6302  &      55  &      3.5  \\
NGC6445  &      42  &      1.5  \\
NGC6537  &      150 &      4.0  \\
NGC7026  &      37  &      1.8  \\
He2-104  &      125 &      10.0 \\
BI Cru   &      214 &      10.0 \\

\hline
\end{tabular}
\end{table}

\section{Conclusions}

The first and most obvious result is that a sharply changing density
distribution cannot reproduce the observed behavior of the luminosity
with inclination angle. The only models that agree well with the
observations have the smoothly varying structure of the last row of
Figure\,3. We will continue calling this distribution a ``disk''.

The 2-D model copes well with double central stars. By using a hot,
compact star with a cooler giant-like companion, we get good agreement
with the observations. Since we only have a relatively small sample of
random systems the spread in our plots is rather large and a precise
and detailed model cannot be selected on this basis. We do, however,
confirm that the observed behavior as a function of inclination angle
is easily reproducible with simple and physically meaningful and
plausible models of bipolar nebulae.

\end{document}